# Co-adsorption of terrylene and tetrapentacontane on hexagonal boron nitride – towards a two-dimensional Shpolskii matrix

Rong Sun[1,2], Tolibjon Abdurakhmonov[1], Jan-Niklas Heidkamp[1], Julian Schröer[1], Paul Weinbrenner[1,2], Kenji Watanabe[3], Takashi Taniguchi[4], Tobias Korn[1,2], Oliver Kühn[1,2], and Friedemann Reinhard[1,2,] ✉

[1] *Institute of Physics, University of Rostock, 18059 Rostock, Germany*
[2] *Department of Life, Light and Matter, University of Rostock, 18059 Rostock, Germany*
[3] *Research Center for Electronic and Optical Materials, National Institute for Materials Science, 1-1 Namiki, Tsukuba 305-0044, Japan*
[4] *Research Center for Materials Nanoarchitectonics, National Institute for Materials Science, 1-1 Namiki, Tsukuba 305-0044, Japan*

✉Email: friedemann.reinhard@uni-rostock.de

**ABSTRACT:** Organic molecular emitters are promising building blocks for quantum technologies, but their use at two-dimensional interfaces is limited by diffusion, aggregation, and photobleaching. We present work towards constructing an ultrathin two-dimensional Shpolskii-like host-guest matrix on hexagonal boron nitride (hBN) by co-depositing terrylene fluorophores with the long-chain tetrapentacontane in a solution-phase process. We demonstrate that this approach disperses terrylene in a roughly one-nanometer-thin alkane layer without detectable phase separation and allows tuning the emitter density from ensembles to isolated emitters showing characteristic blinking behavior consistent with single-molecule fluorescence. Time-lapse confocal microscopy, autocorrelation analysis, and fluorescence recovery after photobleaching show that suppressed diffusion and stable emitter positions can be achieved in several samples with a tetrapentacontane matrix. Encapsulation with hBN from the top side further improves photostability, with emitters demonstrating long lifetimes despite high photon doses. Successful co-adsorption of isolated emitters and alkane matrix is a milestone towards extending Shpolskii-type molecular isolation to hBN and provides a route towards stable, interface-integrated molecular qubits.

Optically addressable organic molecules are emerging as chemically programmable building blocks for quantum technologies [1–10]. Their optical and spin properties can be tailored through molecular design, while their small size and compatibility with mild fabrication processes make them attractive for nanoscale device integration. For quantum sensing and hybrid quantum architectures, surface-confined molecules are particularly attractive because atomic proximity to the surface can enhance coupling to nearby spins and external fields [11–13]. To make use of this advantage, however, the molecules must remain bright, isolated, and positionally stable at the surface, requiring simultaneous control over three key challenges: photobleaching, aggregation, and lateral diffusion.

Hexagonal boron nitride (hBN) provides an appealing platform for this purpose. Its atomically flat surface, wide band gap, and optical transparency make it a highly suitable interface for molecular emitters [14–16]. Recent demonstrations of hBN-supported pentacene molecular qubits with optically detected magnetic resonance further highlight the potential of hBN as a scaffold for molecular quantum systems [17, 18]. hBN encapsulation has been shown to suppress blinking and photobleaching of emitters [19–21], addressing the photostability issue. In contrast, aggregation and lateral diffusion remain a concern. Deposition of fluorescent molecules on hBN often leads to aggregation [22–24] and molecules still diffuse on the surface of hBN [19, 25, 26]. While preferred adsorption at defect sites [17, 18] or cryogenic cooling can help to alleviate the problems and in favorable cases isolate single molecules [27], the field is presently lacking a generic technique to prevent aggregation and diffusion and reliably create immobilized molecules at a tunable density.

In three-dimensional materials, doping of fluorophores into host-guest matrices is an established method to prevent aggregation and diffusion [28–30]. A prominent example is molecular stabilization in Shpolskii matrices, where aromatic guest molecules are dilutely dispersed in crystalline n-alkane hosts [31–33]. The ordered alkane lattice isolates individual emitters, suppresses aggregation, and reduces environmental fluctuations, enabling narrow optical transitions and long-lived fluorescence. Finding a two-dimensional (2D) host-guest matrix could combine molecular isolation with direct access to hBN-based devices. Conventional Shpolskii matrices, however, are typically bulk or relatively thick crystalline hosts and are therefore poorly suited to interface-integrated molecular quantum systems.

Conceptually, Shpolskii matrices could also exist in two dimensions. Long-chain n-alkanes form ordered physisorbed layers on van der Waals surfaces governed by intermolecular packing and molecule-substrate interactions [34]. In particular, hexacontane on hBN has been shown to form lamellar rows, aligned with the hBN lattice [35]. Atomic force microscopy (AFM) and scanning tunneling microscopy (STM) studies have even demonstrated that these layers can host guest molecules [36–38], but have also observed strong phase separation between the guest and the host or even between different host species themselves [39], leading to aggregation. Whether alkanes can act as a host for dispersed single molecules thus is an open question.

Here, we study the co-deposition of tetrapentacontane ($C_{54}H_{110}$) with terrylene fluorophores, aiming to develop an ultrathin host-guest layer on hBN as a surface-confined molecular emitter platform. In this architecture, the long-chain alkane matrix is intended to act as an ordered interfacial host that disperses terrylene molecules and suppresses aggregation and lateral redistribution on hBN. A top hBN layer is then introduced to encapsulate the molecular layer, further improving photostability. By combining molecular isolation with van der Waals encapsulation, this architecture addresses the key requirements for stable surface molecular emitters: isolation, positional stability, and long-term optical stability. Proving that co-adsorption is possible makes a significant step to extend Shpolskii-type molecular isolation to an hBN-based interface and provides a practical strategy for stabilizing organic molecular emitters in van der Waals quantum devices.

Figure 1(a) illustrates the targeted ultrathin terrylene/$C_{54}H_{110}$ host-guest layer on hBN. To realize this, terrylene and $C_{54}H_{110}$ were co-deposited from toluene solution onto cleaned hBN in a single step (Figure S1). This solution-based method deposits the terrylene guest and alkane host simultaneously and allows the terrylene density to be tuned directly through its concentration in solution. Fluorescence microscopy of the resulting samples reveals two notable details: first, homogeneous co-deposition of alkanes and terrylene is possible. The alkane host does not fully displace terrylene from the hBN surface, despite the fact that the long $C_{54}H_{110}$ host has a much higher adsorption energy [40, 41]. Density functional theory (DFT) calculations rationalizes that this co-adsorption is thermodynamically favourable. Although $C_{54}H_{110}$ binds to hBN more strongly than a terrylene molecule (Adsorption energies: -3.97 and -1.87 eV respectively), the competition for a certain surface area is governed by the adsorption energy per unit area. The two species are nearly degenerate with 12.9 vs 13.6 meV/Å$^2$ for alkane and terrylene, respectively. Displacing an adsorbed terrylene by host matrix alkane costs ~95 meV, which is approximately 3.7 $k_BT$ at room temperature. Therefore, the alkane matrix immobilizes the embedded dye molecules rather than expelling them (see Section S6-S7 in Electronic Supplementary Material (ESM) for details). Second, there is no detectable phase separation into terrylene-rich and alkane-rich domains, indicating that terrylene remained well dispersed within the alkane matrix. With the $C_{54}H_{110}$ concentration fixed at 11 µM, samples prepared using 3-, 100-, 1000-, and $10^5$-fold dilutions of a saturated terrylene solution showed a progressive transition from dense, extended fluorescence features to well-separated diffraction-limited spots (Figures 1b,d,f,h,S2,S3). The isolated spots exhibited the characteristic fluorescence spectrum of terrylene (Figure S4). The intensity trajectory of the diffraction-limited spot marked in Figure 1(h) showed abrupt on-off transitions consistent with single-molecule fluorescence (Figure 1g), indicating the isolation of individual terrylene emitters. AFM results further revealed a uniform adsorbate layer, with scratch measurements yielding a thickness of approximately 1 nm (Figures 1c,e,S5), corresponding to only one or a few molecular layers. These results establish that single-step solution co-deposition indeed produces well-dispersed terrylene emitters on hBN.

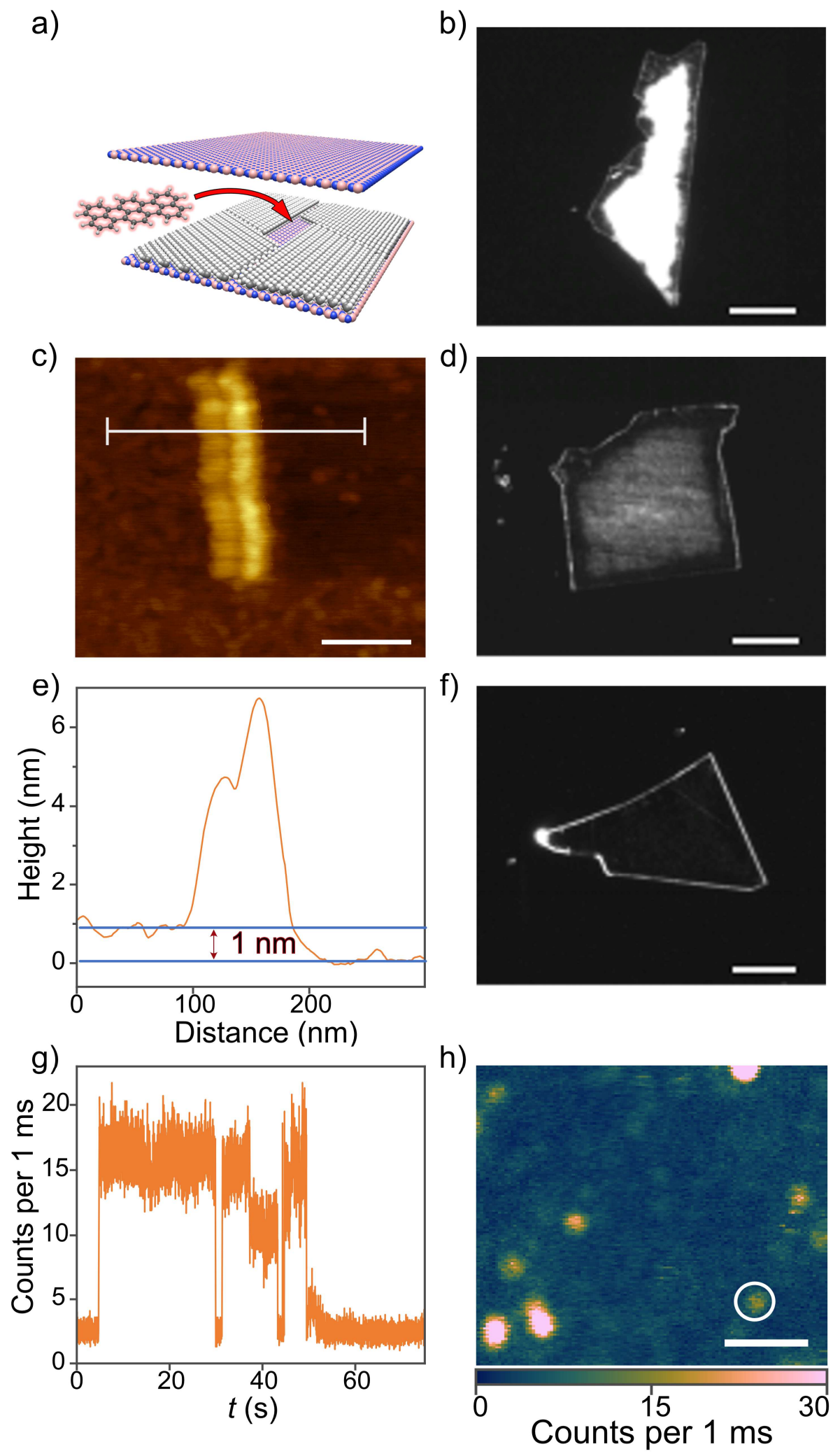

**Figure 1** Formation and characterization of terrylene/$C_{54}H_{110}$ molecular layers on hBN (a) Schematic illustration of the terrylene (orange) / $C_{54}H_{110}$ (gray) molecular layer on hBN (purple) and hBN encapsulation. (b,d,f) Wide-field fluorescence microscopy images of terrylene/$C_{54}H_{110}$ samples prepared with different terrylene concentrations while keeping the $C_{54}H_{110}$ concentration fixed at 11 µM. The terrylene concentrations correspond to 3-, 100-, and 1000-fold dilutions of a saturated terrylene solution, respectively. Scale bars: 10 µm. (c) Atomic force microscopy image of a scratched terrylene/$C_{54}H_{110}$ layer on hBN. Scale bar: 100 nm. (e) Height profile taken along the line in (c), showing an apparent step height of approximately 1 nm. (g) Fluorescence intensity trajectory of the diffraction-limited spot marked in (h), showing abrupt on-off transitions consistent with single-molecule fluorescence and irreversible photobleaching after around 50 s. (h) Confocal fluorescence image of a terrylene/$C_{54}H_{110}$ sample prepared using a $10^5$-fold dilution of the saturated terrylene solution (532 nm laser excitation with an excitation power of 8 µW), showing well-separated diffraction-limited spots. Scale bar: 2 µm.

We next investigated lateral diffusion of terrylene molecules on hBN with and without the alkane matrix. As a reference, time-lapse confocal images of terrylene deposited directly on hBN were acquired under continuous 532 nm excitation. The fluorescence pattern changed markedly over time, with localized fluorescence features appearing and disappearing at different surface positions over seconds to minutes (Figure 2a). This spatial redistribution of the fluorescence pattern indicates that terrylene is not fully immobilized on bare hBN but moves between adsorption sites. Nevertheless, the persistence of prominent fluorescence features between consecutive images suggests that spatial redistribution is much slower than expected on an ideally homogeneous hBN surface. Molecular dynamics simulations of a single terrylene on pristine hBN yielded a diffusion coefficient of $1.12\times10^{-4}$ $cm^2s^{-1}$ (Figures 2e,f). On average the molecule is flat-lying and physisorbed during the whole simulation time. The hBN surface is highly corrugated with the lowest barrier for lateral displacement of 195 meV. Rapid lateral diffusion becomes possible by thermal out-of-plane fluctuations [42]. At this diffusion rate, a molecule would on average be displaced by ~200 µm in a second, orders of magnitude farther than the displacement of the emitters, which remain fixed on a timescale of minutes in the experiment.

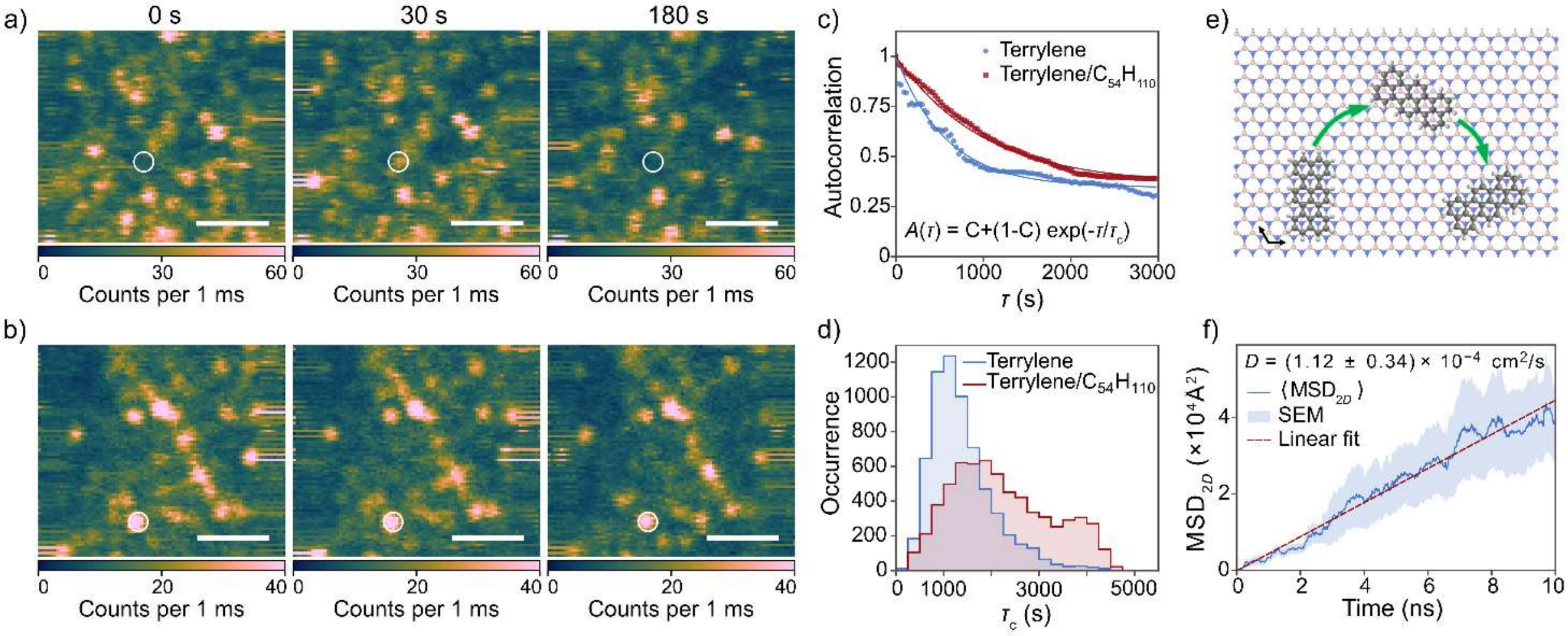

**Figure 2** Short-timescale diffusion of terrylene on hBN. (a) Time-lapse confocal fluorescence images of terrylene deposited directly on hBN under 532 nm laser excitation with an excitation power of 4 μW. The marked position illustrates the appearance and disappearance of fluorescence features over time, indicating lateral redistribution of terrylene on the hBN surface. Scale bars: 2 μm. (b) Time-lapse confocal fluorescence images of terrylene/$C_{54}H_{110}$ on hBN acquired under the same excitation. The fluorescence pattern remains more stable over time, indicating suppressed molecular redistribution in the alkane matrix. Scale bars: 2 μm. (c) Representative normalized temporal autocorrelation functions extracted from the marked pixels in (a) and (b). The solid lines represent the fitting results. (d) Distributions of pixel-wise characteristic decorrelation time ($\tau_c$) for terrylene and terrylene/$C_{54}H_{110}$ samples, showing longer decorrelation times after alkane co-deposition. (e) MD simulation for a single terrylene molecule diffusing on pristine hBN at 300 K, showing that the molecule moves freely across the surface in a flat-lying orientation. Mean-squared displacement (MSD) of the terrylene center of mass from MD simulation. A linear fit shows the Einstein relation $MSD = 4Dt$. The shaded area shows the standard error of the mean (SEM) across individual MD trajectories (details in Section S8)

Real hBN presents a more heterogeneous adsorption landscape than the ideal simulated surface. Atomic defects, wrinkles, step edges, and adsorbed contaminants can change the molecule-surface interaction and create sites with different binding strengths. Terrylene can therefore remain trapped at one site for an extended period before moving to another, resulting in slow, intermittent, and spatially confined diffusion rather than continuous free diffusion. As shown in Figure 1(f), enhanced fluorescence was observed along the edges of the hBN flakes, indicating preferential accumulation of terrylene in these regions. Previous studies likewise found that terrylene preferentially localizes near hBN step edges and wrinkles [27], suggesting that molecules diffuse across the basal plane before becoming trapped at these surface features. The difference between experiment and simulation shows that the intrinsic terrylene-hBN interaction alone cannot account for the observed diffusion dynamics, highlighting the important role of local variations in the surface adsorption potential.

We then examined whether $C_{54}H_{110}$ could modify this adsorption landscape and further reduce terrylene motion. Compared with terrylene on bare hBN, terrylene/$C_{54}H_{110}$ retains a substantially more stable fluorescence pattern during continuous imaging (Figure 2b). Emitters remain at the same locations for longer periods, indicating that the alkane layer increases molecular sticking and suppresses lateral diffusion. To quantify this stabilization over the full field of view, we calculated the temporal autocorrelation function of every pixel from the confocal image sequences. The pixel autocorrelation measures how closely the local fluorescence intensity at a later time resembles its initial value. A stationary fluorescence pattern therefore retains a high correlation, whereas molecular motion redistributes the intensity between pixels and causes faster decorrelation.

To quantify this observation, each single-pixel autocorrelation trace was further fitted using single-exponential decay with formula (1):

$$A(\tau) = C + (1 - C)e^{-\tau/\tau_c} \qquad (1)$$

where $C$ represents the long-time residual correlation and $\tau_c$ is the characteristic decorrelation time. (Figure 2c). Alkane co-deposition shifted the entire $\tau_c$ distribution toward longer times (Figure 2d), increasing its median from 1242 to 2052 s. This approximately 65% increase shows that the alkane matrix prolongs the residence of terrylene at individual surface locations and suppresses its lateral diffusion across hBN. The resulting longer $\tau_c$ (2052 s) is likely limited by photostability rather than diffusion, as no molecular diffusion is visible in the time-lapse microscopy (Figure 2b).

A similar picture of restricted diffusion emerges at even longer timescales of hours to days (Figure 3). To examine molecular redistribution over these timescales, we performed fluorescence recovery after photobleaching (FRAP) measurements. A large field of view was first imaged under weak 532 nm excitation (~ 4 μW). A smaller central region was then photobleached by a higher laser power (~ 2 mW), followed by imaging of the original field of view at defined time intervals. Freshly prepared samples of pure terrylene typically show pronounced and spatially heterogeneous recovery. Localized fluorescence features re-emerged within the bleached region in an irregular and nonuniform manner (Figures 3a,c), indicating redistribution of terrylene across the hBN surface. In contrast, we observed several terrylene/$C_{54}H_{110}$ samples where the photobleached region retains its original shape and remains spatially homogeneous over the same period (Figures 3b,d). This persistence of the bleach pattern suggests that the alkane matrix restricts the lateral transport of terrylene. This pattern is even more clearly visible in one-dimensional line-cuts across the photobleached area (Figures 3e,f). Note that these images present an average over all vertical and horizontal lines crossing the photo-bleached region (see section S5 in ESM).

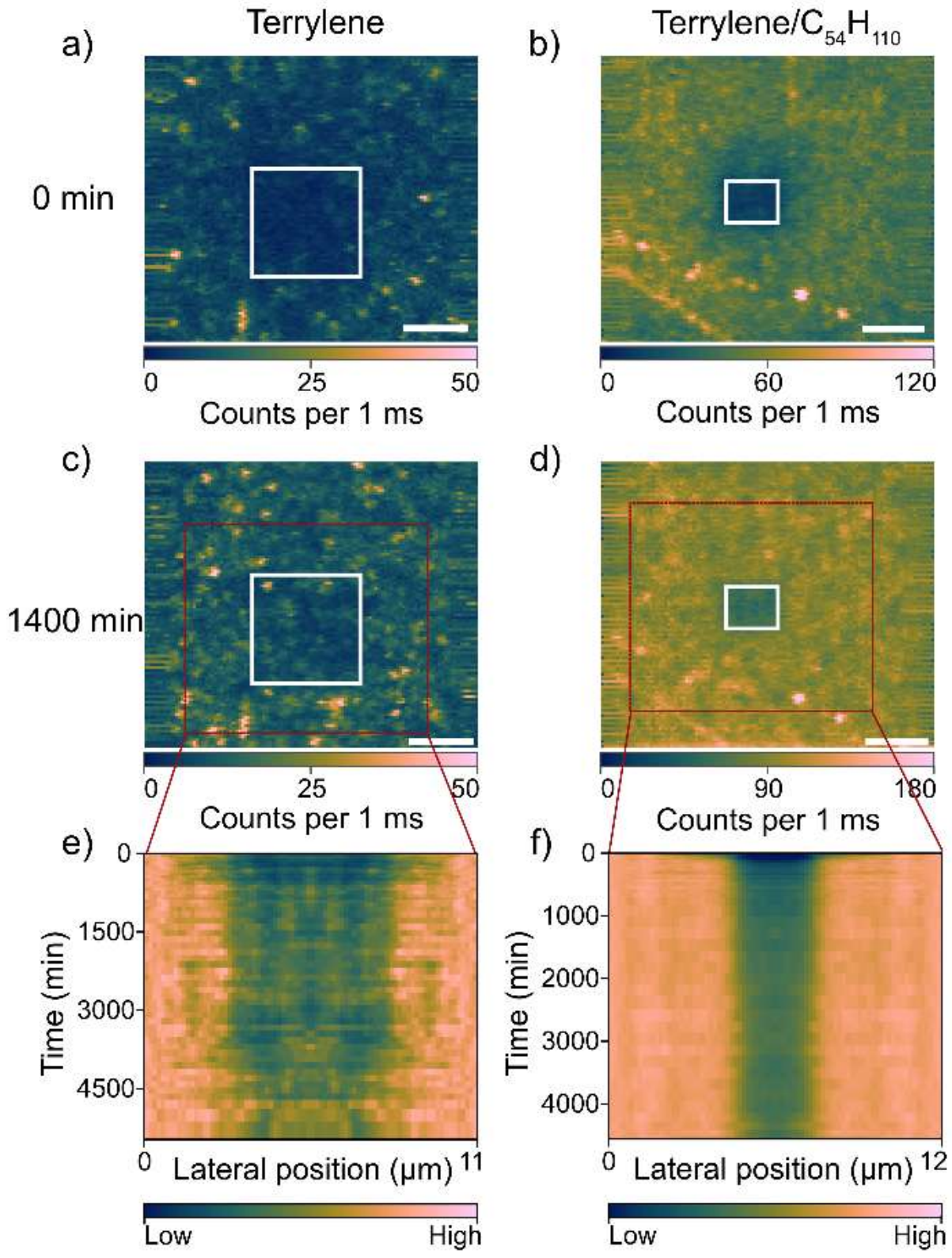

**Figure 3** Long-timescale diffusion of terrylene on hBN. (a,b) Confocal fluorescence images of terrylene on hBN and terrylene/$C_{54}H_{110}$ on hBN immediately after photobleaching, respectively. The white boxes mark the bleached regions. (c,d) Confocal fluorescence images of the same regions after 1400 min of recovery. The terrylene sample shows spatially heterogeneous fluorescence recovery within the bleached region, whereas the terrylene/$C_{54}H_{110}$ sample retains a more well-defined bleached area. The red boxes mark the cropped regions. Scale bars: 3 μm. (e,f) Time-dependent line-cuts through the photo-bleached square extracted from the FRAP image sequences of terrylene and terrylene/$C_{54}H_{110}$ samples, respectively. The line-cuts were generated by averaging the fluorescence intensity data from all four edges of the photobleached region (see section S5 for details). The terrylene sample shows irregular recovery within the bleached region, while the terrylene/$C_{54}H_{110}$ sample maintains a persistent intensity minimum with little internal spatial structure.

We note that an extended FRAP study over a wider range of samples revealed several outlier samples in both directions (Figures S7–S9). We sometimes observe immobilization on bare terrylene samples, especially if they had been exposed to the ambient atmosphere for several days before measurement. We attribute this to formation of a natural alkane matrix from the ambient air that hinders diffusion for long timescales [43]. We also observed residual diffusion on some co-adsorbed terrylene/$C_{54}H_{110}$ samples, especially ones that displayed an unusually low level of terrylene fluorescence hinting towards incomplete deposition. We attribute this either to a second population of mobile fluorophores, potentially terrylene molecules diffusing on top of the alkane layer as have been studied in earlier work [27], or to incomplete deposition that leads to diffusion between islands of alkanes. Nevertheless, Figure 2 and Figure 3 establish a central result of our work. Full immobilization can be achieved in samples prepared by co-adsorption of terrylene with alkanes.

We next explored whether the photochemical stability of the co-adsorbed terrylene emitters could be further improved by hBN encapsulation. A top hBN flake was transferred onto the terrylene/$C_{54}H_{110}$ layer, forming an hBN/terrylene-$C_{54}H_{110}$/hBN heterostructure in which the molecular layer is confined between two hBN surfaces and separated from the ambient environment (Figure 4a). Encapsulated and unencapsulated terrylene/$C_{54}H_{110}$ samples were then continuously imaged under strong 532 nm excitation, and the fluorescence decay was analyzed as a function of accumulated photon dose.

Encapsulation slowed the overall fluorescence loss (Figure 4b). To quantify this statement, we define the "lethal doses" LD50 and LD80 as the accumulated photon doses required to reduce the fluorescence intensity by 50% and 80% from its initial value, respectively. After hBN encapsulation, LD50 increased by approximately a factor of 2, whereas LD80 increased by approximately a factor of 13. The much larger increase in LD80 indicates that encapsulation not only delays the average fluorescence decay but also preserves a long-lived emitter population that remains fluorescent under substantially higher photon doses. Pixel-resolved LD50 maps show this effect spatially: the encapsulated sample contains regions that withstand much larger photon exposure before reaching half of their initial fluorescence, presumably regions of particularly tight encapsulation, whereas the unencapsulated sample bleaches more uniformly and at lower photon doses (Figures 4c,d). We further analyzed the accumulated photon dose received by each pixel at the point of photobleaching (Figure 4e). This bleaching-dose distribution provides a direct measure of the high-dose survival population. Although the encapsulated sample was imaged for 1.5 times longer than the unencapsulated sample, all pixels in the unencapsulated sample photobleached within the experimental window, whereas approximately 3% of pixels in the encapsulated sample remained fluorescent at the end of the measurement. Among the pixels that photobleached, encapsulation strongly enhanced the high-dose tail of the distribution. The fraction of pixels surviving beyond 1.0 × $10^{25}$ photons per $cm^2$ increased from 0.42% to 18.6%. These results show that hBN encapsulation produces a long-lived emitter population with substantially improved resistance to photobleaching.

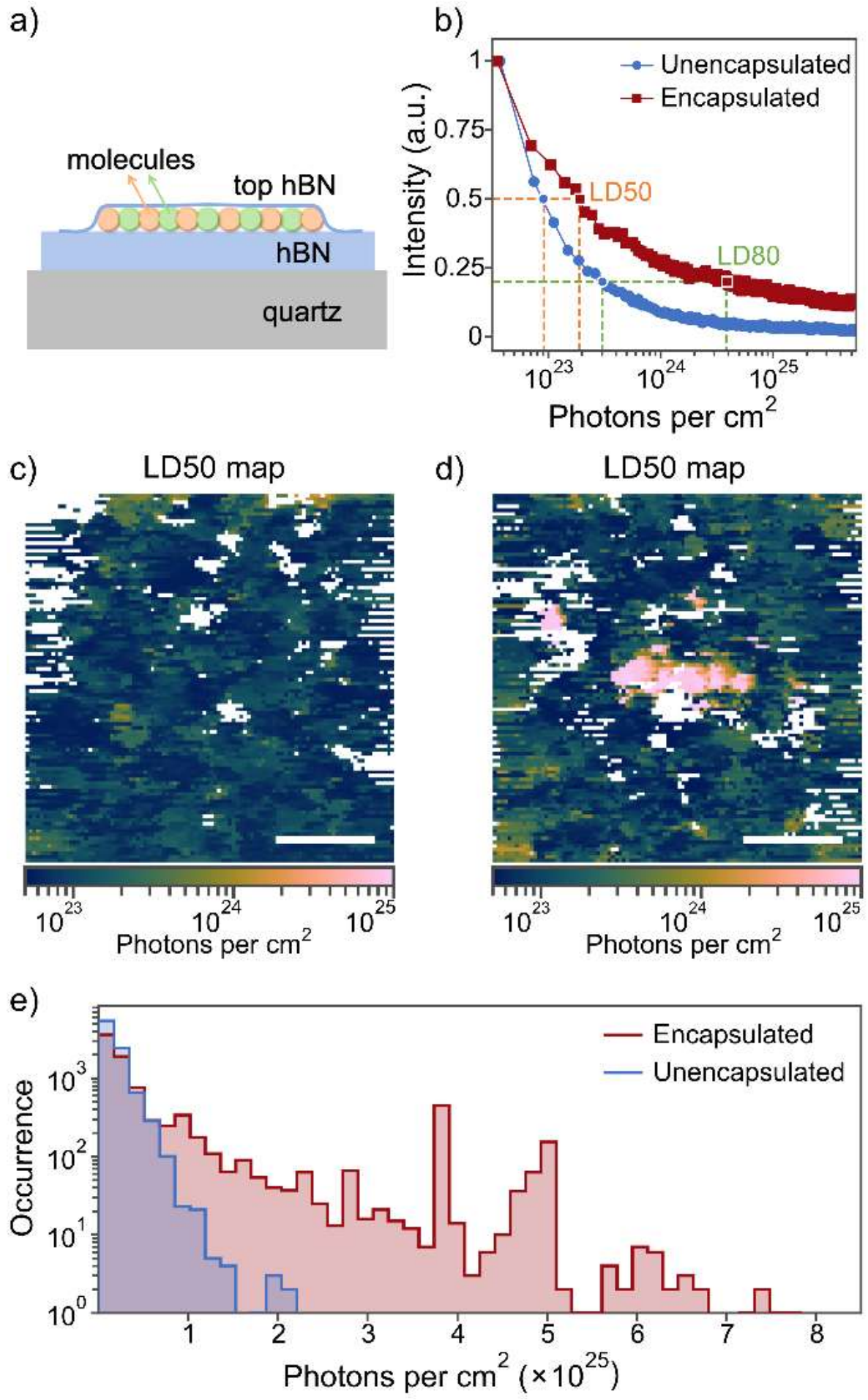

**Figure 4** Enhanced photostability of terrylene/$C_{54}H_{110}$ molecular layers through hBN encapsulation. (a) Schematic illustration of the hBN-encapsulated terrylene/$C_{54}H_{110}$ heterostructure. A top hBN flake covers the terrylene/$C_{54}H_{110}$ layer supported on bottom hBN. (b) Average fluorescence decay curves of encapsulated and unencapsulated terrylene/$C_{54}H_{110}$ samples under continuous 532 nm excitation with a laser power of 233 μW, plotted as a function of accumulated photon dose. (c,d) Pixel-resolved LD50 maps of unencapsulated and encapsulated terrylene/$C_{54}H_{110}$ samples, respectively. The encapsulated sample contains regions that withstand higher photon doses before reaching half of their initial fluorescence. Scale bars: 2 μm. (e) Pixel-wise distribution of photobleaching doses for encapsulated and unencapsulated terrylene/$C_{54}H_{110}$ samples. The photobleaching dose is defined as the accumulated photon dose received when each pixel photobleaches. Pixels that did not photobleach within the experimental window are excluded from the histogram.

The enhanced fluorescence persistence after hBN encapsulation is consistent with previous single-molecule studies. The hBN-covered 6,13-bis(triisopropylsilylethynyl)pentacene (TIPS-pentacene) molecules on glass were reported to survive for hours, whereas TIPS-pentacene molecules on bare glass bleached within about hundreds of seconds [19]. Separately, terrylene molecules on hBN were shown to exhibit much greater photostability than on $SiO_2$/Si or glass [44], highlighting the favorable optical environment provided by hBN. Together, these studies support a stabilizing role of hBN for surface-supported molecular emitters. Quantifying this effect, however, can be complicated by optical interference in hBN-containing layered structures, where an hBN multilayer can act as a spectral filter to attenuate the intensity of the excitation light. Such interference was beyond the scope of the above studies and may also be relevant to the present comparison, because the addition of a top hBN layer can modify both the excitation at the molecular layer and the fluorescence outcoupling efficiency. Although normalization to the initial fluorescence intensity largely removes a static difference in the detected emission intensity, a change in the local excitation field would affect the actual excitation dose experienced by the molecules. The reported LD50 and LD80 values are therefore operational quantities based on the nominal incident photon dose. Without detailed optical modeling of the individual hBN stacks, the magnitude of their enhancement cannot be attributed quantitatively to the protective effect of encapsulation alone. Nevertheless, irrespective of the possible contribution of optical-stack effects, the encapsulated sample exhibited slower fluorescence decay and a pronounced long-lived emitter population under the applied excitation conditions. In addition, the successful encapsulation is noteworthy: the top hBN flake could be transferred onto the molecular layer without visibly displacing the molecules or inducing pronounced aggregation, while retaining their fluorescence (Figure S12c). This demonstrates that hBN encapsulation provides a viable means of mechanically confining and stabilizing surface-adsorbed molecular emitters. By separating the emitters from ambient oxygen and moisture, encapsulation may reduce photo-oxidative bleaching pathways and provide a more protected local environment.

Together, these results show that interfacial engineering can effectively stabilize molecular emitters on hBN. Single-step co-deposition of terrylene with $C_{54}H_{110}$ forms isolated molecules that appear well-dispersed within a nanometer thin homogeneous layer. Their density can be tuned from molecular ensembles to isolated single emitters. Time-lapse confocal imaging, pixel-wise autocorrelation analysis, and FRAP measurements show that several samples with a $C_{54}H_{110}$ matrix display suppressed lateral motion and stabilized emitter positions. Subsequent hBN encapsulation addresses photobleaching as a second instability channel by improving resistance to high photon doses and producing a long-lived emitter population. Future studies will have to unravel the exact conformation of the co-adsorbed layers, in particular address the question whether terrylene and alkane molecules indeed form a host-guest system with well-defined insertion sites, a conclusion which cannot yet be drawn with certainty from the observation of successful co-adsorption, restricted diffusion and improved photostability. For example, polarization-resolved Raman spectroscopy could provide valuable complementary information on the molecular orientation and ordering of the matrix, thereby helping to further constrain the structural picture of the co-adsorbed layer. Ultimately, two-dimensional host-guest matrices promise to provide a materials design strategy for integrating chemically defined organic molecules with two-dimensional platforms, offering a feasible route toward stable, interface-integrated organic molecular qubits.

**Electronic Supplementary Material:** Supplementary material (Sample preparation and wide-field fluorescence microscopy, confocal fluorescence microscopy, atomic force microscopy (AFM), short-timescale diffusion analysis, long-timescale diffusion analysis, adsorption registry of n-alkane, adsorption energy density and chain-length extrapolation, molecular dynamics simulations, encapsulation of terrylene/$C_{54}H_{110}$ on hBN and references) is available.

# Acknowledgements

This work is funded by the Deutsche Forschungsgemeinschaft (DFG, German Research Foundation)—SFB 1477 ‘Light–Matter Interactions at Interfaces’, project no. 441234705 and IRTG 2676/1-2023 ‘Imaging of Quantum Systems’, project no. 437567992. T. K. and J. H. acknowledge financial support via DFG project KO3612/8-1 (project No. 549364913). K.W. and T.T. acknowledge support from the CREST (JPMJCR24A5), JST and World Premier International Research Center Initiative (WPI), MEXT, Japan.

# Declaration of competing interest

All the contributing authors report no conflict of interests in this work.

# Author contribution statement

R. S.: Data curation, formal analysis, investigation, methodology, visualization, writing – original draft. T. A.: Data curation, formal analysis, methodology, visualization, writing – original draft. J. H.: Methodology. J. S.: Methodology. P. W.: Methodology, Writing – review & editing. K. W.: Resources. T. T.: Resources. T. K.: Funding acquisition, supervision, writing – review & editing. O. K.: Methodology, supervision, writing – review & editing. F. R.: Conceptualization, funding acquisition, project administration, resources, supervision, writing – review & editing. All the authors have approved the final manuscript.